\documentclass{revtex4}
\usepackage{amsmath}
\usepackage{amssymb}
\usepackage{graphicx}

\begin{document}

\title{Physical singularity in the regular spacetime and fundamental length}

\author{Vladimir Dzhunushaliev}

\affiliation{Department of Physics. and Microelectronics Engineering, 
Kyrgyz-Russian Slavic University, Bishkek, Kievskaya Str. 
44, 720021, Kyrgyz Republic,\\
dzhun@krsu.edu.kg}

\author{Ratbay Myrzakulov}

\affiliation{Institute of Physics and Technology, 480082, Almaty-82, Kazakhstan\\
cnlpmyra@satsun.sci.kz}

\begin{abstract}
It is shown that formally regular solutions in 5D Kaluza-Klein gravity have singularities.  This phenomenon is connected with the existence of a minimal length in nature. The  calculation of the derivative of the $G_{55}$ metric component leads to the appearance 
of the Dirac's $\delta-$function. In this case the Ricci scalar becomes singular since  there is a square of this derivative. 
\end{abstract}

\maketitle

\section{Introduction}
Practically any fundamental physical theory has singularities and 
one of the main aim of the modern physics is the struggle against 
these singularities. For example, in the classical electrodynamics 
there is the divergence of the electric field near to a point-like 
electric charge; in quantum electrodynamics there are divergences 
connecting with loops in Feynman diagrams. In general relativity 
there are cosmological, Schwarzschild, Reissner-Nordstr\"om and so 
on singularities (for review see, for example, 
Ref.\cite{Rendall:2005nf}). Most of these singularities are 
connected with a punctual origin of a field (electric, gravitational 
and so on). Probably the most general receipt against singularities 
is to extend the point-like particle in some directions. String 
theory do it by extension the particle in 1 dimension that leads to 
the replacement of 0-dimensional particle to 1-dimensional string.
\par
Intuitively we understand what is it the singularity: a place where 
some invariants are divergent, for example, the field strength, 
scalar curvature and so on. These singularities are mathematical in 
the sense that they are present in the solution of the corresponding 
field equations. In this notice we would like to show that there 
exist such situations when the mathematical solution is regular but 
from the physical point of view there is a physical (soft) 
singularity. In the case presented here it is connected with the 
fact that in nature there is a minimal length (Planck length). Some 
physical quantities vary quickly during the Planck length, for 
example, from +1 to -1. In this case this quantity is similar to the 
step (Hevisaide) function. The derivative of such function is the 
Dirac's $\delta-$function. It is not bad but if some invariant has 
such derivative in a square then it is bad: we will have a 
\textit{physical} singularity. In other words such \textit{physical 
singularity is present in a mathematical regular solution only if 
some physical quantity vary too quickly during of minimal length. } 

\section{Regular wormhole-like flux tube solution}

In this notice we would like to show that in 5D Kaluza-Klein gravity 
there are regular solutions but in some points there are the 
conditions for the appearance of a physical singularuty: $G_{55}$ 
metric component change too quickly. At first we will present the 
regular 5D wormhole-like flux tube metric
\begin{eqnarray}
    ds^2 & = & \frac{dt^{2}}{\Delta(r)} - \Delta(r) e^{2\psi (r)}
    \left [d\chi + \omega (r)dt + Q \cos \theta d\varphi \right ]^2
    \nonumber \\
    &-& dr^{2} - a(r)(d\theta ^{2} +
    \sin ^{2}\theta d\varphi ^2),
\label{sec1-10}
\end{eqnarray}
where $\chi $ is the 5$^{th}$ extra coordinate; $r,\theta ,\varphi$ 
are $3D$ spherical-polar coordinates; $r \in \{ - \infty, + \infty 
\}$ is the longitudinal coordinate; $Q$ is the magnetic charge.
\par
On the 4D language we have the following components of the 4D 
electromagnetic potential $A_\mu$
\begin{equation}
A_t = \omega (r) \quad \text{and} \quad A_\varphi = Q \cos \theta 
\label{sec1-20}
\end{equation}
and the Maxwell tensor is
\begin{equation}
    F_{rt} = \omega' (r) \quad
    \text{and} \quad
    F_{\theta \varphi} = -Q \sin \theta .
\label{sec1-30}
\end{equation}
Substituting this ansatz into the 5-dimensional Einstein vacuum 
equations
\begin{equation}
    R_{AB} - \frac{1}{2} \eta_{AB} R = 0
\label{sec1-35}
\end{equation}
$A,B = 0,1,2,3,5$ and $\eta_{AB}$ is the metric signature, gives us
\begin{eqnarray}
    \frac{\Delta ''}{\Delta} - \frac{{\Delta '}^2}{\Delta^2} +
    \frac{\Delta 'a'}{\Delta a} + \frac{\Delta ' \psi '}{\Delta} +
    \frac{q^2}{a^2 \Delta ^2}e^{-4 \psi} & = & 0,
\label{sec1-40}\\
    \frac{a''}{a} + \frac{a'\psi '}{a} - \frac{2}{a} +
    \frac{Q^2}{a^2} \Delta e^{2\psi} & = & 0,
\label{sec1-50}\\
    \psi '' + {\psi '}^2 + \frac{a'\psi '}{a} -
    \frac{Q^2}{2a^2} \Delta e^{2\psi} & = & 0,
    \label{sec1-60}\\
    - \frac{{\Delta '}^2}{\Delta^2} + \frac{{a'}^2}{a^2} -
    2 \frac{\Delta ' \psi '}{\Delta} - \frac{4}{a} +
    4 \frac{a' \psi '}{a} +
    \frac{q^2}{a^2 \Delta ^2} e^{-4 \psi} +
    \frac{Q^2}{a^2} \Delta e^{2\psi} & = & 0,
\label{sec1-70}\\
          \omega ' &=& \frac{q}{a \Delta ^2} e^{-3 \psi} .
\label{sec1-70a}
\end{eqnarray}
$q$ is the electric charge. As the electric $q$ and magnetic $Q$ 
charges are varied it was found \cite{dzhsin1} that the solutions to 
the metric in Eq. \eqref{sec1-40}-\eqref{sec1-70} evolve in the 
following way :
\begin{enumerate}
\item
$0 \leq Q < q$. The solution is \emph{a regular gravitational flux 
tube}. The solution is filled with both electric and magnetic 
fields. The longitudinal distance between the $\pm r_H$ surfaces 
increases, and the cross-sectional size does not increase as rapidly 
as $r \rightarrow r_H$ with $q \rightarrow Q$. The values $r= \pm 
r_H$ are defined by the following way $\Delta(\pm r_H)=0$.
\item
$Q = q$. In this case the solution is \emph{an infinite flux tube} 
filled with constant electric and magnetic fields. The 
cross-sectional size of this solution is constant ($ a= const.$).
\item
$0 \leq q < Q$. In this case we have \emph{a singular gravitational 
flux tube} located between two (+) and (-) electric and magnetic 
charges at $\pm r_{sing}$. At $r = \pm r_{sing}$ this solution has 
real singularities where the charges are placed.
\end{enumerate}
We will consider the case with $q \approx Q$ but $q > Q$. In this 
case there is a region (throat) $|r| \leq r_H$ where the solution is 
like to a tube filled with almost equal electric and magnetic 
fields. The length $L = 2 r_H$ of the throat depends on relation 
$\delta = 1 - Q/q$ in the following manner $L \stackrel{\delta 
\rightarrow 0}{\longrightarrow} \infty$ but $\delta > 0$. The 
numerical analysis \cite{dzh3} shows that the spatial cross section 
of the tube $a(\pm r_H) \approx 2 a(0)$. The cross section of the 
tube at the center $a(0)$ is arbitrary and we choose it as $a(0) 
\approx l^2_{Pl}$. This allows us to say that we have a super-thin 
and super-long gravitational flux tube: $L \gg \sqrt{a(0)}$, $a(r) 
\approx a(0) \approx l_{Pl}, \left| r \right| \leq r_H$.

\begin{figure}[h]
  \begin{center}
    \fbox{
    \includegraphics[height=7cm,width=7cm]{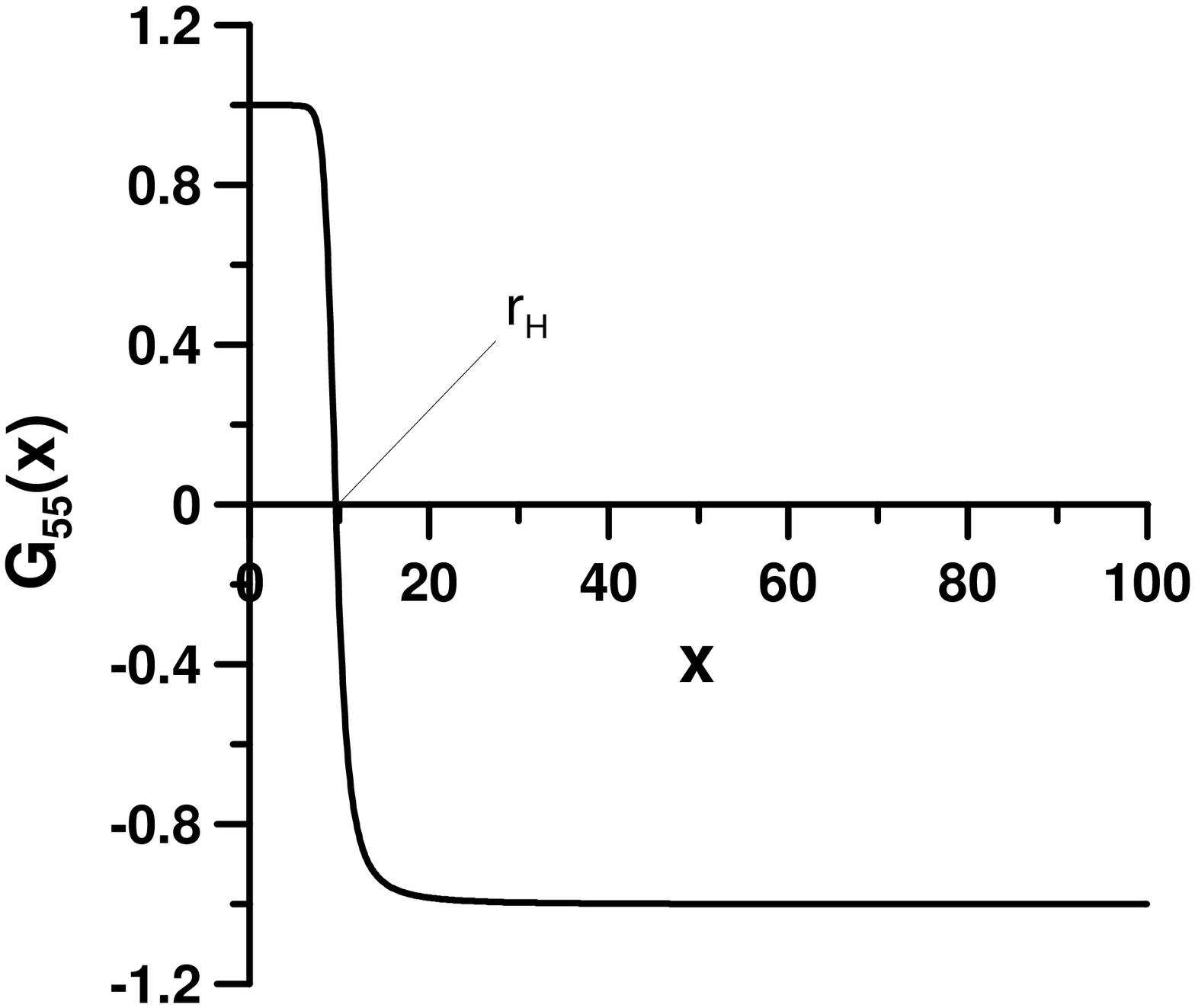}}
    \caption{The metric component $G_{55}(x)$,
    $x = r/\sqrt{a(0)}$ is the dimensionless longitudinal coordinate.}
    \label{fig:prod}
  \end{center}
\end{figure}
\par
In Fig. \eqref{fig:prod} the profile of the $G_{55} = \Delta \mathrm 
e^{2\psi}$ metric component is presented \cite{dzh3}. We see that 
nearby the value $r = r_H$ this function changes drastically from 
the value $G_{55} \approx +1$ by $r \lesssim + r_H - l_0$ to $G_{55} 
\approx -1$ by $r \gtrsim + r_H + l_0$. The same occurs near $r = - 
r_H$.
\par
On the throat by $\left| r \right| < r_H - l_0$ the solution 
approximately is (since $q \approx Q$)
\begin{eqnarray}
    a(r) & \approx & \frac{Q_0^2}{2} = const,
\label{sec1-160}\\
    e^{2\psi(r)} & \approx & \frac{1}{\Delta} = \cosh^2\frac{r}{\sqrt{a(0)}},
\label{sec1-170}\\
    \omega(r) & \approx & \sqrt{2}\sinh\frac{r}{\sqrt{a(0)}} ,
\label{sec1-180}\\
    G_{55}(r) & = & \Delta \mathrm e^{2 \psi(x)} \approx 1 .
\label{sec1-190}
\end{eqnarray}
Such approximation is valid only by $|r| \lesssim r_H - l_0$ where 
$l_0$ is some small quantity $l_0 \ll r_H$.
\par
Now we would like to estimate the length $l_0$ of region where the 
change of the metric component $G_{55} = \Delta \mathrm e^{2\psi}$ 
is
\begin{equation}
    \left. \Delta \mathrm e^{2\psi}\right|_{r \approx r_H + l_0} -
    \left. \Delta \mathrm e^{2\psi}\right|_{r \approx r_H - l_0} \approx 2 .
\label{sec1-195}
\end{equation}
For this estimation Eq. \eqref{sec1-50} will be used. On the throat 
by $\left| r \right| < r_H - l_0$ this equation approximately is
\begin{equation}
    - \frac{2}{a} + \frac{Q^2}{a^2} \Delta e^{2\psi} \approx 0.
\label{sec1-200}
\end{equation}
We can estimate $l_0$ by solving Einsten's equations 
\eqref{sec1-40}-\eqref{sec1-70} nearby $r = + r_H$ (for $r = - r_H$ 
the analysis is the same) and define $r = r_H - l_0$ where the last 
two terms in Eq. \eqref{sec1-50} have the same order
\begin{equation}
    \left.
        \left(
            \frac{2}{ a}
        \right)
    \right|_{r=r_H - l_0} \approx
    \left.
        \left(
            \frac{Q^2}{a^2} \Delta e^{2\psi}
        \right)
    \right|_{r=r_H - l_0}
\label{sec1-210}
\end{equation}
The solution close to $r = + r_H$ we search in the following form
\begin{equation}
    \Delta(r) = \Delta_1 \left( r_H - r \right) +
    \Delta_2 \left( r_H - r \right)^2 + \cdots .
\label{sec1-230}
\end{equation}
The substitution in Eq. \eqref{sec1-40} gives us the following 
solution
\begin{equation}
    \Delta_1 = \frac{q e^{-2\psi_H}}{a_H}.
\label{sec1-260}
\end{equation}
After the substitution into Eq. \eqref{sec1-210} we have
\begin{equation}
    l_0 \approx \sqrt{a(0)} = l_{Pl}
\label{sec1-270}
\end{equation}
here we took into account that the numerical analysis \cite{dzh3} 
shows that $a_H = a(\pm r_H) \approx 2 a(0)$. It means that the 
change of macroscopical dimensionless function $G_{55} = \Delta 
\mathrm e^{2\psi}$ as in Eq. \eqref{sec1-195} occurs during the 
Planck length. The metric \eqref{sec1-10} by $|r| \approx r_H - 
l_{Pl}$ approximately is
\begin{equation}
    ds^2 \approx \mathrm e^{2\psi_H} dt^2 - dr^2 - a(r_H)
    \left( d\theta^2 + \sin \theta d \varphi^2 \right) -
    \left(
        d\chi + \omega dt + Q \cos \theta d \phi
    \right)^2
\label{sec1-273}
\end{equation}
by $|r| \approx r_H + l_{Pl}$ the metric approximately is
\begin{equation}
    ds^2 \approx -\mathrm e^{2\psi_H} dt^2 - dr^2 - a(r_H)
    \left( d\theta^2 + \sin \theta d \varphi^2 \right) +
    \left(
        d\chi + \omega dt + Q \cos \theta d \phi
    \right)^2
\label{sec1-276}
\end{equation}
here we took into account that numerical calculations \cite{dzh3} 
show that $\psi \approx \psi_H = \mathrm {const}$ by $|r| \gtrsim 
r_H$. We see that during the Planck length the metric signature 
changes from $\{ +,-,-,-,- \}$ to $\{ -,-,-,-,+ \}$. Simultaneously 
it is necessary to mention that the metric \eqref{sec1-10} is 
non-singular by $|r| = r_H$ and approximately is \cite{dzh3}
\begin{eqnarray}
    ds^2 \approx &&
    e^{2 \psi_H} dt^2 -
    e^{\psi_H} dt \left( d\chi + Q \cos\theta d \varphi \right) -
    dr^2 - a(r_H) \left( d\theta^2 + \sin^2 \theta d\varphi^2 \right) = 
\nonumber \\
    && \left[
        e^{\psi_H} dt - \frac{1}{2} \left( d \chi + Q \cos \theta d \varphi \right)
    \right]^2 -
    dr^2 - a(r_H) \left( d\theta^2 + \sin^2 \theta d\varphi^2 \right) -
\nonumber \\
    &&    
    \frac{1}{4} \left( d \chi + Q \cos \theta d \varphi \right)^2
\label{sec1-280}
\end{eqnarray}
where $\psi_H$ is some constant.
\par
If we write the metric \eqref{sec1-10} in the 5-bein formalism
\begin{equation}
    ds^2 = \omega^A \omega^B \eta_{AB} , \quad 
    \omega^A = e^A_\mu dx^\mu , \quad
    x^\mu = t,r,\theta , \varphi , \chi
\label{sec1-290}
\end{equation}
then we see that
\begin{eqnarray}
  \eta_{AB} &=& \left\{ +1,-1,-1,-1,-1 \right\} \quad
  \text{by} \quad |r| \lesssim r_H - l_{Pl}
\label{sec1-300}\\
  \eta_{AB} &=& \left\{ -1,-1,-1,-1,+1 \right\} \quad
  \text{by} \quad |r| \gtrsim r_H + l_{Pl}
\label{sec1-310}
\end{eqnarray}

\section{Physical singularity}

Let us consider more carefully the situation at the place $r = \pm 
r_H$. For this more it is convenient to present the metric 
\eqref{sec1-10} in the form
\begin{equation}
    ds^2 = \frac{e^{2\psi (r)}}{\tilde{\Delta}(r)} dt^2 - \tilde{\Delta}(r)
    \left [d\chi + \omega (r)dt + Q \cos \theta d\varphi \right ]^2
    - dr^{2} - a(r)(d\theta ^{2} +
    \sin ^{2}\theta d\varphi ^2),
\label{sec1-360}
\end{equation}
where
\begin{equation}
    G_{55} = \tilde{\Delta}(r) = \Delta(r) e^{2\psi (r)} .
\label{sec1-370}
\end{equation}
By the definition we have
\begin{equation}
    \left.\tilde{\Delta}'(r) \right|_{r = r_H} =
    \lim \limits_{\Delta r \rightarrow 0}
    \frac{\tilde{\Delta} \left( r_H + \Delta r \right) -
    \tilde{\Delta} \left( r_H - \Delta r \right)}{2 \Delta r} .
\label{sec1-380}
\end{equation}
Now we have to take into account the assumption that in the nature 
there is a minimal length (Planck length). In this case one can see 
that here there is one subtlety. \textit{The point is that $\Delta 
r$ can not converge to zero since there is the minimal length}. 
Strictly speaking the introduction of the minimal length can be made 
in the quantum gravity only. Now we have not such theory 
nevertheless there is some algebraic consideration which allows us 
to understand how the Planck length can appear. 
Ordinary the existence of the minimal length does not result in 
problems at the calculation of a derivative but not in this case. 
Equations \eqref{sec1-195} and \eqref{sec1-380} give us
\begin{eqnarray}
    \left.\tilde{\Delta}'(r) \right|_{r = r_H} \approx
    &&
    \frac{\tilde{\Delta} \left( r_H + l_{Pl} \right) -
    \tilde{\Delta} \left( r_H - l_{Pl} \right)}{2 l_{Pl}} =
    \frac{\Theta \left( r_H + l_{Pl} \right) -
    \Theta \left( r_H - l_{Pl} \right)}{2 l_{Pl}} \approx
\nonumber \\
    &&
    \left.\Theta'(r) \right|_{r = r_H}
    \approx \left. \delta \left(r - r_H \right) \right|_{r = r_H} =
    \delta \left( 0 \right) = \infty
\label{sec1-390}
\end{eqnarray}
where
\begin{equation}
    \Theta \left( x \right) =
    \left\{
    \begin{array} {rl}
        +1, & \mbox{if } x > 0 \\
        0, & \mbox{if } x = 0 \\
        -1, & \mbox{if } x < 0
    \end{array}
    \right.
\label{sec1-400}
\end{equation}
is the step function, $\delta(r)$ is the Dirac's delta-function. 
More exactly
\begin{equation}
    \tilde{\Delta}' \left( \pm r_H \right) \approx
    \frac{1}{l_{Pl}} .
\label{sec1-405}
\end{equation}
It means that near to $r = \pm r_H$
\begin{equation}
    \tilde{\Delta}'(r) \approx \delta \left( r \right).
\label{sec1-410}
\end{equation}
It is not bad if we have only $\tilde{\Delta}'(r)$ not $\tilde 
\Delta '^2 (r)$. But in our case the situation is bad since, for 
example, the Ricci scalar is
\begin{eqnarray}
    \frac{1}{2} R(r) = && \frac {1}{2} \left(
        \frac {\tilde \Delta'^2(r)}{{\tilde{\Delta}}^{2} (r)} -
        \frac {q^2 e^{-4 \psi (r)} }{a^2(r) \tilde{\Delta}^2 (r)}
    \right)
    + {\psi'}^{2}(r) -
    \frac {\psi'(r) {\tilde{\Delta}'}(r)}{\Delta (r)} +
    2 \psi'' (r) +
\nonumber \\
    && 2 \frac {a'(r) \psi' (r)}{a(r)} -
    \frac {1}{2} \frac {{a'}^2 (r)}{a^2 (r)} +
    2 \frac {a'' (r)}{a(r)} -
    \frac {2}{a (r)} +
    \frac {1}{2} \frac {Q^2 \tilde{\Delta} (r)}{a^2 (r)}
\label{sec1-420}
\end{eqnarray}
At the first sight we have not any problems since
\begin{equation}
    \left.
        \frac {\tilde \Delta'^2(r)}{{\tilde{\Delta}}^{2} (r)} -
        \frac {q^2 e^{-4 \psi (r)} }{a^2(r) \tilde{\Delta}^2 (r)}
    \right|_{r = \pm r_H} = 0
\label{sec1-425}
\end{equation}
in the consequence of Einsteins's equation \eqref{sec1-40}. But this 
is not the whole of history since the above mentioned analysis of 
the derivative $\tilde \Delta'(r)$ in Eq. \eqref{sec1-390} shows us 
that Ricci scalar has $\tilde \Delta'^2(r) = \delta^2(0) = \infty$. 
Approximately we have
\begin{equation}
    R\left( \pm r_H \right) =
    \left.\left(
        \frac {\tilde \Delta'^2}{{\tilde{\Delta}}^{2}} -
        \frac {q^2 e^{-4 \psi} }{a^2 \tilde{\Delta}^2}
    \right) \right|_{r = \pm r_H} \approx
    \frac{1}{l^2_{Pl}} \frac{1}{\tilde \Delta (\pm r_H)} =
    \frac{1}{l^2_{Pl}} \frac{1}{0}
\label{sec1-427}
\end{equation}
It means that by $r = \pm r_H$ we have a singularity - \textit{a 
soft} singularity. The word soft signifies that in this case 
singularity differs from the \textit{hard} singularity of the 
Schwarzschild black hole. One can call such singularity a 
\textit{quantum} singularity since it is connected with such quantum 
gravity phenomenon as the minimal length.

\section{Conclusions}

In this paper we have shown that such quantum gravity phenomenon as 
the minimal length leads to the very interesting physical 
consequences: the appearance of a physical singularity. The reason 
of such singularity is that there is some physical quantity which 
changes too quickly during the Planck length. Formal mathematical 
calculations of the derivative of this physical quantity do not show 
any problems but the presence of the minimal length leads to the 
fact that the corresponding derivative at this point is the Dirac's 
$\delta-$function. Finally the Ricci scalar has this derivative in a 
square and it is a pure infinity (singularity). This analysis is 
correct only in the case when the cross section of the gravitational 
flux tube is $\approx l^2_{Pl}$, if this quantity $\gg l^2_{Pl}$ 
then we will have not any problems.
\par
Also it is interesting to note the in Ref's \cite{guendelman}, \cite{Zaslavskii:2004bk} the 4D analog 
(Levi-Civita-Robimson-Bertotti metric \cite{Levi-Civita}) of the 5D 
gravitational flux tube solutions is considered as a model of the 
electric charge.


\begin{thebibliography}{99}

\bibitem{Rendall:2005nf}
A.~D.~Rendall, ``The nature of spacetime singularities,'' 
gr-qc/0503112.

\bibitem{dzhsin1}
V. Dzhunushaliev and D. Singleton, \textit{Phys. Rev.} \textbf{D59}, 064018 
(1999).

\bibitem{dzh3}
V.~Dzhunushaliev, ``Strings and Branes under Microscope'', to be published in Annalen der Physik, gr-qc/0312038; \\
``Wormhole solutions in 5D Kaluza-Klein theory as string-like objects,''
New Developments in Quantum Cosmology Research. 
Series: Horizons in World Physicss, Ed. A. Reimer, volume 247, 113-140 (2005), 
(New-York, Nova Science Publishers, 2005), 
gr-qc/0405017.

\bibitem{guendelman}
E. I. Guendelman, \textit{Gen. Relat. Grav.}, \textbf{23}, 1415 (1991).

\bibitem{Zaslavskii:2004bk}
O.~B.~Zaslavskii,
\textit{Phys. Rev.}, {\bf D70}, 104017 (2004);\\
``N-spheres in general relativity: Regular black holes without apparent
horizons, static wormholes with event horizons and gravastars with a tube-like core,''
gr-qc/0601017.

\bibitem{Levi-Civita}
T. Levi-Civita, \textit{Atti Acad. Naz. Lincei}, \textbf{26}, 519 (1917) ; \\
B. Bertotti, \textit{Phys. Rev.}, \textbf{116}, 1331 (1959); \\
I. Robinson, \textit{Bull. Akad. Pol.}, \textbf{7}, 351 (1959).


\end{thebibliography}
\end{document}